# Efficient World-line-based Quantum Monte Carlo Method Without Hubbard-Stratonovich Transformation

J. Wang[a], W. Pan[a] and D. Y. Sun[a,b*]

[a]*Department of Physics, East China Normal University, 200241 Shanghai, China*
[b]*Shanghai Qi Zhi Institute, Shanghai 200030, China*

## Abstract

By precisely writing down the matrix element of the local Boltzmann operator ($e^{-\tau h}$, where $h$ is the Hermitian conjugate pairs of off-diagonal operators), we have proposed a new path integral formulation for quantum field theory and developed a corresponding Monte Carlo algorithm. With current formula, the Hubbard–Stratonovich transformation is not necessary, and is not based on the determinant approach, which can improve the computational efficiency. The results show that, the simulation time has the square-law scaling with system sizes, which is comparable with the usual first-principle calculation. The current formula also improves the accuracy of the Suzuki–Trotter decomposition. As an example, we have studied the one-dimensional half-filled Hubbard model at finite temperature. The obtained results are in excellent agreement with the known solutions. The new formula and Monte Carlo algorithm can be used in various studies.

[*]Corresponding author.
E-mail address: dysun@phy.ecnu.edu.cn (D.-Y. Sun).

# I. Introduction

Strongly correlated quantum many-body (SCQMB) systems possess a rich physical phenomenon, making them an essential topic in condensed matter physics. To overcome an exponentially increased dimensions of the Hilbert space, as well as the intrinsically strong correlation of SCQMB systems, scientists have strongly advocated for efficient and accurate numerical methods.[1-6] Over the last decades, several numerical methods have been proposed, such as the exact diagonalization (ED) method,[7] the density matrix renormalization group (DMRG) method,[8-10] as well as various quantum Monte Carlo (QMC) methods.[11-13] These numerical methods have played an important role in improving the understanding of SCQMB systems over the past few decades.

In ED methods, the Hamiltonian matrix of systems is directly diagonalized using advanced mathematical techniques, but they are limited to relatively small systems. The DMRG method shares some similarities with the ED methods but it can handle larger systems, making it a more effective method for low-dimensional systems. Some examples of commonly used QMC methods include determinant QMC (DQMC),[11,14,15] auxiliary field QMC (AFQMC),[12,16,17] and diagrammatic MC (DiagMC).[13,18] Recently, we have developed a new method to directly calculate the ground state properties of elements of the Hamiltonian matrix, for a class of special systems.[19,20]

The AFQMC method can be considered as a typical example of the DQMC method. In both DQMC and AFQMC methods, the quartic fermion operators are decomposed into a quadratic form by performing the Hubbard–Stratonovich transformation.[21-23] For the DQMC method[11], a number of important advancements[24,25] and improved algorithms have been developed.[14] The DQMC method has been used to determine physical properties at zero[6,26,27] and finite temperatures.[26,28] The typical AFQMC method is a projection technique in which the operator $e^{-\beta H}$ continuously acts on a trial wave function; accordingly, the ground state properties[12,17,24,27] can be evaluated. To avoid the fermion sign problem or phase problem, AFQMC is usually implemented by an advanced restriction on the sample paths,[29] and recently, the AFQMC has additionally been developed for finite temperatures.[16,30-32] DiagMC[13] combines the MC technique with Feynman diagrams of the perturbative expansion to calculate physical quantities,[18,33-37] and several generations[38-41] and improvements[34,37,38,42,43] of this method have been developed based on the original DiagMC. For a more comprehensive introduction to various numerical methods, refer to review articles, such as those in Ref.[2,3]

In parallel to determinantal formulation mentioned above, the world-line formulation represents another catalog of QMC.[44-48] In the world-line QMC, the direct-space and imaginary-time is used to interpolate the representation of the fermion fields. The advantage of the world-line QMC is avoiding the time-consuming process of evaluating fermion determinants. However, in the world-line QMC, a closed path with non-zero weight is not always easy to sample in many-body wavefunction (WF) spaces. At present, several successful examples seem to be limited to a few specific Hamiltonians, and a universal algorithm is waiting to propose.

To further reduce the gap between the experimental studies and QMC calculations, more efficient numerical methods are needed to meet the current demands for studying SCQMB systems. Thus, there is an urgent need to either develop new numerical methods or optimize known ones. In this paper, we propose a new world-line QMC method by introducing a representation of the path integral formula in quantum field theory. This method can be used to calculate various properties of a system at finite temperature. The new formulation does not require the Hubbard–Stratonovich transformation, and not require determinants, therefore the calculation time can be significantly reduced. The results of the test calculation on the one-dimensional Hubbard model are in excellent agreement with exact values.[49,50]

## II. Proposed Formula

To illustrate our method, we chose the simple and representative Hubbard model as an example. It is worth noting that our method can be directly extended to any model or Hamiltonian. The Hamiltonian of the Hubbard model reads:

$$H = -t \sum_{i,j,\sigma}(c_{i\sigma}^\dagger c_{j\sigma} + H.c.) + U \sum_i n_{i\uparrow} n_{i\downarrow}, \qquad (1)$$

where $c_{i\sigma}^+(c_{i\sigma})$ denotes the creation (annihilation) of an electron with spin $\sigma = \uparrow, \downarrow$ at the $i$-th lattice site. The first term on the right-hand side of Eq. (1) represents the one-body term, which accounts for the hopping of electrons between different sites, and $t > 0$ is the hopping amplitude. The second term is the two-body on-site Coulomb interaction, where $U$ represents the interaction strength, and $n_{i\sigma} = c_{i\sigma}^\dagger c_{i\sigma}$ is the number operator for spin $\sigma$ at the site $i$. For convenience, the spin index is omitted in the following description; yet it is included later on to prevent confusion.

One of the key step of our method is to combine each off-diagonal term in the Hamiltonian and its Hermite conjugate into pairs, namely $h_{ij} = -t(c_i^\dagger c_j + c_j^\dagger c_i)$. Clearly $h_{ij}$ is therefore a Hermitian operator. In the case of a general Hamiltonian, $h_{ij}$ can be made up of the pair of a quartic fermion operator and its Hermitian conjugation. The purpose of this combination is to make $h_{ij}$ as a Hermitian operator, and its eigenfunction be easily obtained.

A many-body WF in the occupation number representation is labeled as $|ijK\rangle$. Here, the occupancy of the site $i$ and $j$ is explicitly given, while the occupancy of the rest of the sites is represented by K. For the site $i$ and $j$, the occupation has four cases, which are labeled as $|i\bar{j}K\rangle, |\bar{i}jK\rangle, |ijK\rangle,$ and $|\bar{i}\bar{j}K\rangle$. Here, $\bar{i}(\bar{j})$ indicates that there is no electron occupying the site $i$ $(j)$ site, while $i$ $(j)$ indicates an electron occupying the site $i$ $(j)$ site.

It is easy to prove that $h_{ij}$ has only two eigenstates with non-zero eigenvalues:

$$|\varphi_{ij}\rangle_+ = \tfrac{1}{\sqrt{2}}(|i\bar{j}K\rangle + |\bar{i}jK\rangle) \ ; \ |\varphi_{ij}\rangle_- = \tfrac{1}{\sqrt{2}}(|i\bar{j}K\rangle - |\bar{i}jK\rangle),$$

where the eigenvalues are equal to $-\theta t$ and $\theta t$, respectively. $\theta$ is the sign produced by the particle exchange as $h_{ij}$ acts on $|i\bar{j}K\rangle$. When an even number of exchanges occur, $\theta = 1$, otherwise, $\theta = -1$. The remaining WFs orthogonal to $|\varphi_{ij}\rangle_+$ and $|\varphi_{ij}\rangle_-$ are also the eigenstates of $h_{ij}$, however, the corresponding eigenvalues are zero.

It is easy to demonstrate that in the particle number representation, the non-zero matrix elements of the local Boltzmann operator (LBO), $e^{-\tau h_{ij}}$, where $\tau$ can be any number, are only present in the following cases:

$$\langle i\bar{j}K|e^{-\tau h_{ij}}|\bar{i}jK'\rangle = \langle \bar{i}jK|e^{-\tau h_{ij}}|i\bar{j}K'\rangle = \frac{1}{2}\delta_{K,K'}\left(e^{\theta\tau t}-e^{-\theta\tau t}\right), \quad (2.1)$$

$$\langle i\bar{j}K|e^{-\tau h_{ij}}|i\bar{j}K'\rangle = \langle \bar{i}jK|e^{-\tau h_{ij}}|\bar{i}jK'\rangle = \frac{1}{2}\delta_{K,K'}\left(e^{\theta\tau t}+e^{-\theta\tau t}\right), \quad (2.2)$$

$$\langle ijK|e^{-\tau h_{ij}}|ijK'\rangle = \langle \bar{ij}K|e^{-\tau h_{ij}}|\bar{ij}K'\rangle = \delta_{K,K'}, \quad (2.3)$$

where the remaining matrix elements of $e^{-\tau h_{ij}}$ are equal to zero. Since the operator $e^{-\tau U n_{i\uparrow}n_{i\downarrow}}$ is diagonal, the non-zero matrix element reads:

$$\langle i_\uparrow i_\downarrow K|e^{-\tau U n_{i\uparrow}n_{i\downarrow}}|i_\uparrow i_\downarrow K'\rangle = \delta_{K,K'}e^{-\tau U}. \quad (2.4)$$

Since $t > 0$, the matrix elements in Eqs. (2.2 – 2.3) are always positive, regardless of the value of $\theta$. The matrix element in Eq. (2.1) is negative if $\theta$ is negative, otherwise, it is positive. Eq. (2.1) produces the off-diagonal scattering in WF for the site $i$ and $j$, but Eqs. (2.2 – 2.3) are the diagonal scattering in WF.

The partition function of a quantum many-body system is expressed as $Z = Tr[e^{-\beta H}]$, where $\beta$ is the inverse temperature (or imaginary time). According to the standard path integral formula, the imaginary time ($\beta$) is divided into $m$ time slices, where the partition function then becomes $Z = Tr[(e^{-\tau H})^m]$, with the time step $\tau = \beta/m$. Using the Suzuki–Trotter decomposition,[21-23] the operator $e^{-\tau H}$ can be further decomposed as

$$e^{-\tau H} = \prod_{ij} e^{-\tau h_{ij}} \prod_i e^{-\tau U n_{i\uparrow}n_{i\downarrow}}.$$

Finally, the partition function reads:

$$Z = Tr \prod_{ij} e^{-\tau h_{ij}} \prod_i e^{-\tau U n_{i\uparrow}n_{i\downarrow}} \cdots \prod_{ij} e^{-\tau h_{ij}} \prod_i e^{-\tau U n_{i\uparrow}n_{i\downarrow}} = \sum_\omega \rho(\omega), \quad (3)$$

To calculate the partition function, a complete set of states are inserted between LBOs. For the purpose of our discussion, we number each LBO ($e^{-\tau h}$, $h = h_{ij}$ or $U n_{i\uparrow}n_{i\downarrow}$) in Eq. (3) from right to left as, 1st, 2nd, …$i$-th LBO. The WF following the $k$-th LBO is then denoted by the $k$-th WF. A closed world line (a closed WF sequence, or a closed path) in the WF space is labeled as $\omega = \{\cdots|s\rangle\cdots\}$, where $|s\rangle$ is the $s$-th WF. $\rho(\omega)$

is the associated Boltzmann weight of the world line $\omega$.

Because only the matrix elements in Eqs. (2.1 – 2.4) are non-zero, for any closed world line, the Boltzmann weight has the following form:

$$\rho(\omega) = \left(\frac{1}{2}(e^{\tau t\theta}+e^{-\tau t\theta})\right)^{n_+} \left(\frac{1}{2}(e^{\tau t\theta}-e^{-\tau t\theta})\right)^{n_-} (e^{-\tau U})^{n_0}, \qquad (4)$$

where $n_-, n_+, n_0$ correspond to the number of occurrences of the matrix elements in Eq. (2.1, 2.2, and 2.4) for a closed world line $\omega$, respectively.

Eq. (4) is the key formula of our newly proposed method, which represents the weighting of a world line $\omega$ in the path integral formula. Eq. (4) shows that our method has several novel features: 1) The new formula does not include the Hubbard–Stratonovich transformation, and thus it does not require the auxiliary field; 2) Our algorithm is not based on the determinant approach, thus the heavy calculation related to determinants is absent; 3) Our formula improves the accuracy of Suzuki–Trotter decomposition. According to the matrix identity $e^{-\tau(A+B)} = e^{-\tau A}e^{-\tau B}e^{0.5\tau^2[A,B]}$ with $[A,B] = AB - BA$, the error induced by the Suzuki-Trotter decomposition ($e^{-\tau(A+B)} \approx e^{-\tau A}e^{-\tau B}$) is directly related to $e^{0.5\tau^2[A,B]}$. According to $E = -\frac{1}{Z}\frac{\partial Z}{\partial \beta}$, the error in energies can be estimated by $-\left\langle \frac{\partial e^{0.5\tau^2[A,B]}}{e^{0.5\tau^2[A,B]}\partial \tau}\right\rangle = -\tau\langle AB - BA\rangle$, where $\langle \cdots \rangle$ refers the ensemble average. If $A$ and $B$ are Hermitian conjugate, then $[A,B] \equiv 1$, which always contributes to the energy with an amount of $-\tau$. If $A$ and $B$ are not Hermitian conjugate, as a first-order approximation, $\langle AB - BA\rangle \approx \langle A\rangle\langle B\rangle - \langle B\rangle\langle A\rangle = 0$. Thus, we have enough reason to believe that, when $A$ and $B$ are Hermitian conjugate, $\langle AB - BA\rangle$ contributes the largest error in Suzuki-Trotter decomposition. Since in our method, the Hamiltonian is decomposed into Hermitian conjugate pairs, this error is automatically disappeared.

### III. New Monte Carlo Algorithm

To implement the new formula presented in Eq. 3-4 into the QMC simulations, we have subsequently developed an efficient algorithm. Although it is easy to calculate the weighting of each path, it is relatively difficult to find a closed world line with a non-zero weight, due to the fact that the weight of most paths is zero.

The choice of each WF is very significant for obtaining a closed world line with a non-zero weighting. Here, we design an algorithm similar to the world-line algorithm[51] and the multiple time threading algorithm.[52] The current QMC algorithm contains two steps. To illustrate our method, we present an example in Figure 1, in which the 4-site Hubbard model at the half-filled state is shown. Suppose $\omega_o = \{\cdots |s\rangle \cdots\}$ is a closed world line from the last QMC step, the red line in Figure 1 marks $\omega_o$. The first step is to generate an intermediate world line ($\omega'$) from a randomly selected WF from $\omega_o$. Specifically, a WF, say $|r\rangle$, is randomly selected from $\omega_o$. In Figure 1, the randomly chosen WF is marked by the arrow $A$. Then $|r\rangle$ is scattered by the $r$-th LBO ($e^{-\tau h_r}$), and a new WF $|r+1\rangle'$ is generated by $|r+1\rangle' = e^{-\tau h_r}|r\rangle$. Next, the ($r$+2)-th WF $|r+2\rangle'$ is generated by $|r+2\rangle' = e^{-\tau h_{r+1}}|s+1\rangle'$. This process continues until all the LBOs are cycled in the same sequence as in Eq. (3). In Figure 1, this process starts from the arrow $A$ to the right hand.

In the above scattering process, $e^{-\tau U n_{i\uparrow} n_{i\downarrow}}$ does not produce any bifurcation because it is diagonal. However, for the operator $e^{-\tau h_{ij}}$, if one of the $i$-th or $j$-th sites is occupied and the other is empty, the scattering will be bifurcated. One side of the bifurcation corresponds to Eq. (2.1), in which the occupancy of the $i$-th and $j$-th sites is exchanged before and after scattering (corresponding to the diagonal line in Figure 1). While the other path corresponds to Eq. (2.2), in which the wave function is unchanged before and after scattering (corresponding to the horizontal line in Figure 1). For the bifurcation, we use a similar heat-bath algorithm[53] to select the new path, namely, the path following Eq. (2.1) with the probability of $\rho_{HB}(-) = \frac{|e^{\theta\tau t} - e^{-\theta\tau t}|}{2e^{\tau t}}$, and the other path following Eq. (2.2) with the probability of $\rho_{HB}(+) = \frac{(e^{\theta\tau t} + e^{-\theta\tau t})}{2e^{\tau t}}$. Evidently $\rho_{HB}(-) + \rho_{HB}(+) = 1$. Note $\rho_{HB}(\pm)$ is different from that in Ref. [51].

After the scattering process finished, the intermediate world line $\omega' = \{\cdots |s\rangle' \cdots\}$ is successfully generated. The blue line in Figure 1 is the intermediate world line generated by the above scattering process. It needs to point out that, $\omega'$ may not be a closed world line. In fact, in most cases, it is an open world line. In Figure 1, $\omega'$ is opened at the arrow $A$. The key point is that $\omega'$ may have multiple intersections with $\omega_o$. The intersection means at which the WF is identical in both $\omega_o$ and $\omega'$. In Figure 1, the arrow $A$ and $B$ mark the two intersections. In the second step, the fragment between two randomly chosen intersections in $\omega'$ is used to replace the corresponding part in $\omega_o$, then a new closed world line ($\omega_n$) is constructed, illustrating in the lower panel of Figure 1.

The acceptance rate of the new world line is determined by the ratio of two factors, namely $accpt = \frac{\rho(\omega_n)}{\rho(\omega_o)} \cdot \frac{1}{\rho_{HB}(\pm)}$, where $\rho(\omega_n)$ and $\rho(\omega_o)$ is the Boltzmann weight of the new and old world line according to Eq. 4, and $\rho_{HB}(\pm) = \frac{|e^{\theta\tau t} \pm e^{-\theta\tau t}|}{2e^{\tau t}}$ is the additional weight attached to the heat-bath sampling, which should be deducted from the acceptance rate. If the change is accepted, the updated WFs are implemented. Otherwise, the unchanged WFs are implemented.

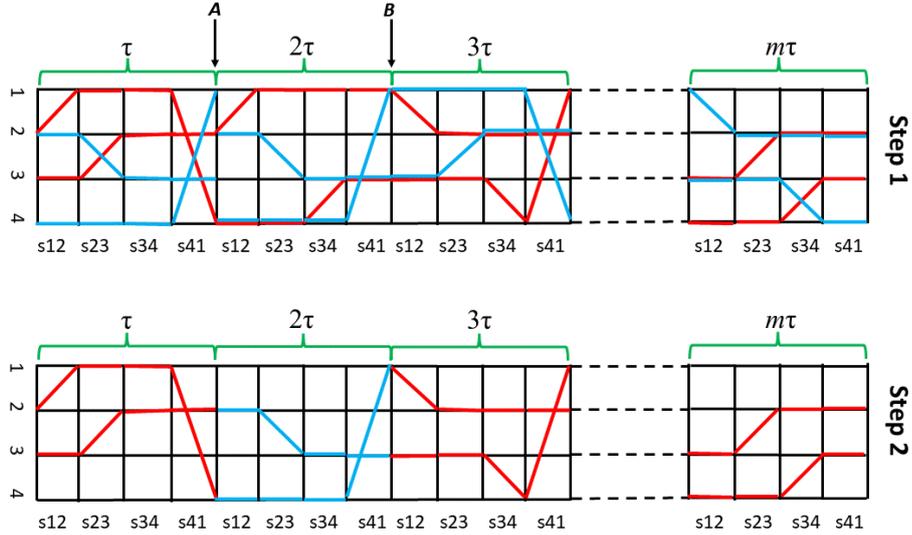

**Figure 1. The graphic representation of the two-step sampling technique for a ring of four sites half-filled Hubbard model. The horizontal and vertical direction refers the time slices and lattice sites, respectively. sij is the shorthand of $e^{-\tau h_{ij}}$, indicating the scatting operator between two wavefunctions (vertical lines). The diagonal operator $e^{-\tau U n_{i\uparrow} n_{i\downarrow}}$ is omitted. In upper panel (step 1), the red line denotes a closed world line ($\omega_o$), while the blue line denotes an open world line**

($\omega'$). **The arrow *A* and *B* mark intersections between $\omega_o$ and $\omega'$. In lower panel (step 2), a closed world line ($\omega_n$) is constructed by replacing the segment between arrows *A* and *B* in $\omega_o$ with that in $\omega'$.**

There are a few differences between the current method and previous ones[45,51]. 1) Except for the initially selected WF $|r\rangle$ from $\omega_o$, the scattering process is irrelevant to the rest WFs in $\omega_o$, which is different from Ref. [51]. 2) In the current method, the sequence of $e^{-\tau h_{ij}}$ in Eq. 3 can be arranged in any way, the only requirement is to combine each off-diagonal term in the Hamiltonian and its Hermite conjugate into pairs; 3) In current method, there are two steps to generate a new closed world line ($\omega_n$). The first step is to generate an intermediate world line ($\omega'$) from a exist closed world line ($\omega_o$) by scattering process. The second step is to construct $\omega_n$ from $\omega'$ and $\omega_o$. The current procedure does not care whether $\omega'$ is closed or not, but $\omega'$ has at least two intersections with $\omega_o$. In previous method[51], the scattering process is aimed to directly generate a closed world line in a single step. Because of this requirement, the previous method[51] usually needs a specific break-up or rearrangement of Hamiltonian. In comparison, the current algorithm is straighter forward, and can be easily extended to any Hamiltonian. 4) As first feeling, one may expect there should be few intersections between $\omega'$ and $\omega_o$. However, the probability of finding a new closed world line using the current method is remarkably high, i.e., close to 100 %. This may contribute the Hermite pairs used in our method. 5) Because the scattering keeps the number of particles unchanged, similar to Ref[51], the current method also work in a canonical ensemble, different from Ref[45].

### IV. Tests on Hubbard Model

The Hubbard model can appropriately describe many favorable and interesting characteristics of correlated electronic systems. Many studies based on the Hubbard model have been carried out to investigate the metal-insulator transition, superconductivity, and magnetic properties caused by electronic correlation. Lieb and

Wu[54] have obtained the exact solution of the ground state for the half-filled one-dimensional Hubbard model. In the last few decades, various theoretical calculations have been carried out to study the one-dimensional Hubbard model. However, few studies have been conducted using this model at finite temperatures.[49,50,55-60]

To illustrate the reliability of our new formula and the new QMC algorithm, we have studied the one-dimensional half-filled Hubbard model at finite temperature. In all calculations, $t$ is taken to have units of energy and is set to $t = 1.0$. The one-dimensional Hubbard model with the number of lattice size of $N = 6, 12$, and 24 have been systematically studied. For each system, the strength of the interaction has also been investigated for $U = 2, 4, 6$ and 8. To determine how the simulation time scaling with $N$, we also calculate a few larger systems with $N$ up to 96. For most simulations, the total number of QMC steps at each temperature or $m$ is more than $10^7$, where the first third of the steps are used to equilibrate the system and the remaining two thirds of the steps are used to calculate the physical properties.

It needs to point out that, like most QMC methods, our new formula could not give a general solution for the sign problem too. However, for particular special cases, the sign problem is not encountered,[26,61] for example, in one-dimensional Hubbard model. By choosing appropriate boundary conditions (periodic or antiperiodic) in one-dimensional Hubbard model, $\theta$ in Eq. (2.1) can be always positive, thus the sign problem can be avoided in current method. This is why we choose the one-dimensional Hubbard model.

The energy, double occupancy, local magnetic moments, and spin correlation functions have been calculated in the temperature range of 0.05 to 4.0. According to thermodynamics, the energy of systems can be calculated as $E = -\frac{1}{Z}\frac{\partial Z}{\partial \beta}$. The double occupancy, denoting the probability of two electrons occupying one site, is written as $O_d = \langle n_{i\uparrow} n_{i\downarrow} \rangle$. $L_0 = \frac{3}{N}\sum_i \langle (S_i^z)^2 \rangle$ is the local magnetic moment, where $S_i^z = \frac{1}{2}\sum_i (n_{i\uparrow} - n_{i\downarrow})$ denotes the $z$-component spin operator at the $i$-th site. The nearest-neighbor and next-nearest-neighbor spin correlation functions are defined as $L_1 =$

$\frac{1}{N}\langle S_i^z S_{i+1}^z \rangle$ and $L_2 = \frac{1}{N}\langle S_i^z S_{i+2}^z \rangle$.

The convergence test on $\tau$ is shown in Fig. 2 for $N=6$, where the upper and lower panels present the data for $T=0.25$ and 0.5, respectively. It can be seen that, with an increase of the number of time slices ($m$), the energy converges quickly. For $T=0.25$, as $m = 80$, corresponding to $\tau = 0.05$, the QMC results approach the exact value with a difference of approximately 1 % and 5 % for $U = 4$ and 8, respectively. Owing to the intrinsic characteristics of the path integral formula, the exact value can be obtained only as $\tau$ approaching to zero. To estimate the QMC energy at $\tau = 0$, the extrapolation to $\tau = 0$ is performed. For this purpose, the data shown in Fig. 2 is fitted by $E = E_0 + a * M^{-c}$, where $E_0$, $a$ and $c$ are fitting parameters. We find the value of $c$ is around 1.5, which is weakly dependent on $U$ and slightly reduces with the increase of $N$. $E_0$ is the extrapolated energy at $\tau = 0$. The exact energy ($E_e$) and extrapolated energy ($E_0$) are shown in Fig. 2. The results show that our QMC method does converge to the exact value at $\tau = 0$.

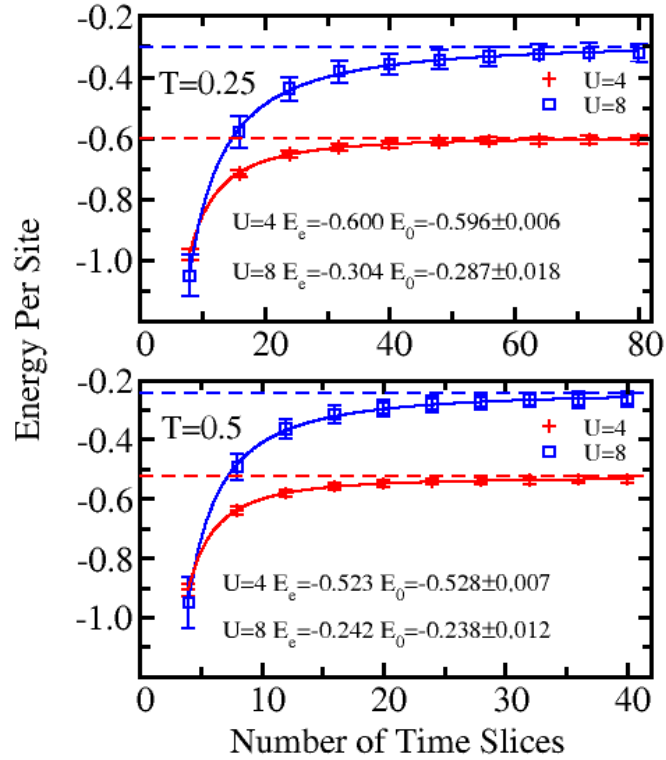

**Figure 2. Energy per site as a function of the number of time slices (*m*) for systems**

with six lattice sites at temperature of 0.25 (upper panel) and 0.5 (lower panel) at both $U = 4$ and 8. The symbol, dashed lines, and solid line represent the QMC data, the exact value, and the fitting curves, respectively. $E_0$ and $E_e$ refer the QMC energy extrapolated at $\tau = 0$ and the exact energy, respectively.

We have tested the convergence speed for several systems with different $N$ and $U$. To comparing the convergence speed, we define a convergence criterion ($\tau^*$), at which the derivative of energy with $m$ is less than 0.0005. This criterion is equivalent to that the change in energy is less than 0.0005 as $m$ increasing by a unit. Fig. 3 shows how $\tau^*$ changing with $U$ and $N$. It can be seen that the convergence speed does not change significantly with $N$, but decreases with the increase of $U$. For most cases, $\tau = 0.05$ is already a good approximation. In the following QMC simulations, $\tau$ is fixed at the value of 0.05.

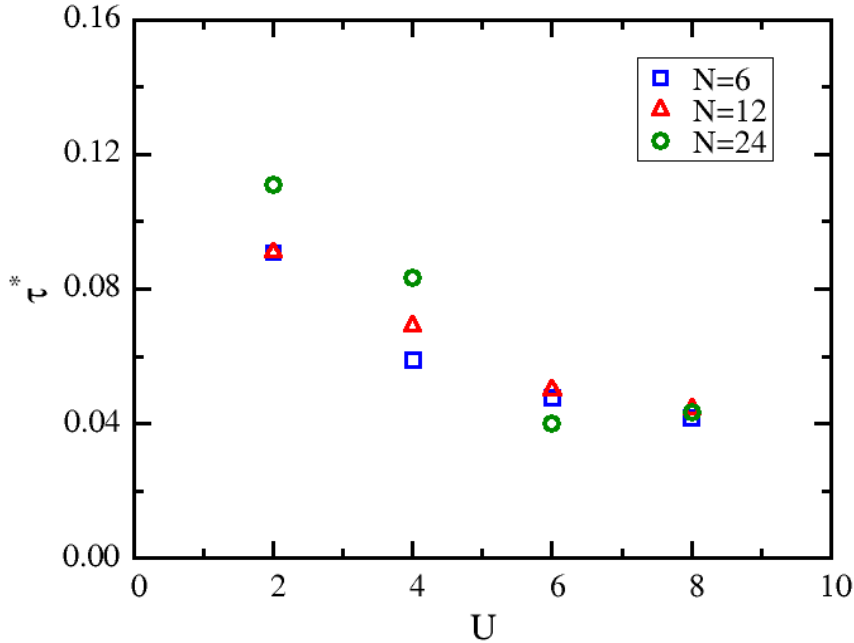

Figure 3. The convergence criterion ($\tau^*$) varying with the on-site Coulomb interaction strength ($U$). Square, triangle and circle symbols represent the systems with lattice size of $N$ = 6, 12 and 24, respectively.

Since our QMC method is determinant independent, the simulation time should have much better scaling with $N$ and $M$. To check this point, we have calculated the simulation time as a function of $N$ and $M$ at fixed $U = 4$. These calculations are performed on a desktop computer with the CPU basic frequency of 3.20GHz (Intel Core I7-8700) and a serial QMC program. The simulation time for 200 thousand MC steps is calculated for various systems, which is summarized in Fig. 4. From this figure, one can see that, the simulation time has the linear scaling with $M$ (upper panel of Fig. 4) and the square-law scaling with $N$ (lower panel of Fig. 4). This computational cost is far lower than other QMC methods involving determinant calculations, and is comparable with the common first-principle calculation. It should be stressed that, our current QMC code could be further improved for the higher efficiency.

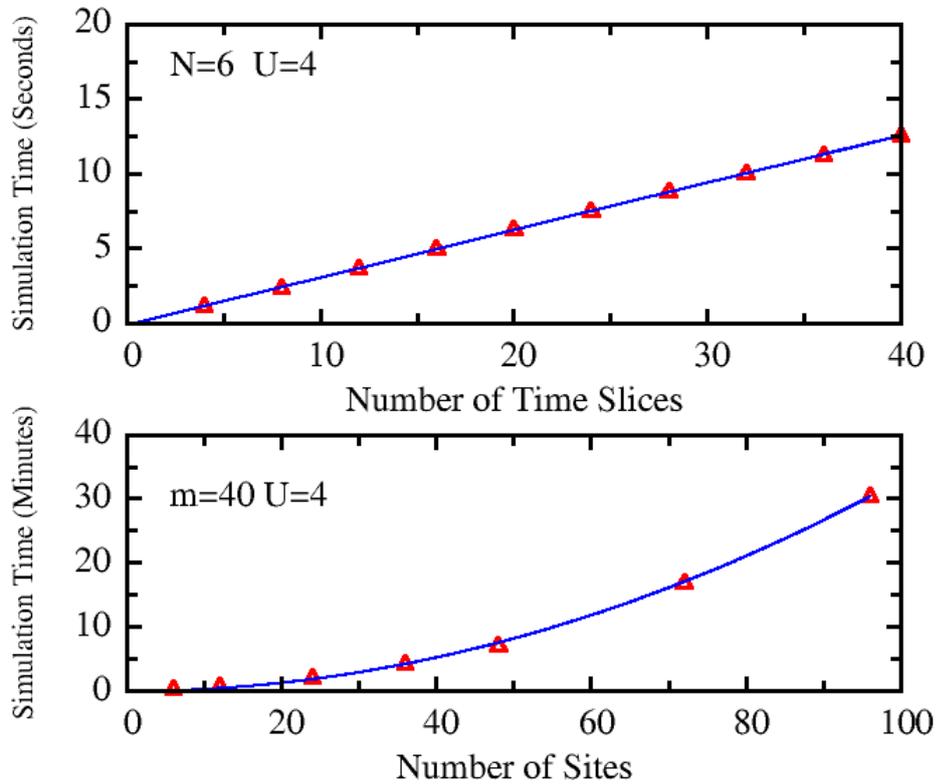

**Figure 4. The simulation time as a function of the number of time slices for N=6 (upper panel) and the number of lattice size for *m*=40 (lower panel). The symbol and solid line represent the QMC data and the fitting curves, respectively. Here**

**the linear fitting to QMC data is adopted in upper panel, while the quadratic function is used to fit QMC data in lower panel.**

In the following, we will present detailed calculation of various physics quantities for the system with six lattice sites. Fig. 5 depicts the energy, double occupancy, and specific heat via temperature for $U = 4$ (left panel) and 8 (right panel) of the system with six lattice sites. The energy calculated by the QMC simulation is in excellent agreement with the exact value for the entire range of temperatures within the error bar (upper panel of Fig. 5). Compared to $U = 4$, there is an evident plateau in the temperature range of 1.0 to 2.0 for $U = 8$. The plateau reflects the fact that the on-site Coulomb interaction has a strong effect in suppressing the occurrence of double occupancy. The change in double occupancy with temperature supports this conclusion.

From the middle panel of Fig. 5, it can be seen that, from the high temperature to the lower temperature, the double occupancy first decreases and then increases for $U = 4$ and $U = 8$. Although the double occupancy increases at low temperatures, the total energy is further reduced with a corresponding decrease in temperature. This reflects the fact that a small increase of the delocalization doublon further decreases the total energy.[49,62] For $U = 4$, the increase in the double occupancy is more evident than when $U = 8$ at lower temperatures. At a fixed temperature, the double occupancy is larger than it is for $U = 8$, demonstrating how the on-site interaction has a noticeable impact on the formation of the double occupations.

The specific heat as a function of temperature is shown in the lower panel of Fig. 5. To calculate the specific heat, the exponential fitting method[63-65] is adopted with the fitting form of $E(T) = E(0) + \sum_{n=1}^{M} c_n e^{\frac{-n\alpha}{T}}$, where $E(0)$, $c_n$ and $\alpha$ are the fitting parameters. In this study, the value of $M$ was 8. From Fig. 5, it can be seen that there are two obvious peaks in the specific heat for $U = 8$. Specifically, there is a narrow peak at low temperatures and a broad peak at high temperatures. In contrast, for $U = 4$ the two peaks become much closer and begin to merge together. The structure of the obtained specific heat peak is consistent with previous findings.[49,57,58] It is believed that this feature in the specific heat is associated with the spin-wave excitations at low temperatures and the single-particle excitations at high temperatures. Thus, spin fluctuations and charge fluctuations are dominant at low and high temperatures, respectively, which can be highlighted by the correlation functions (see below).

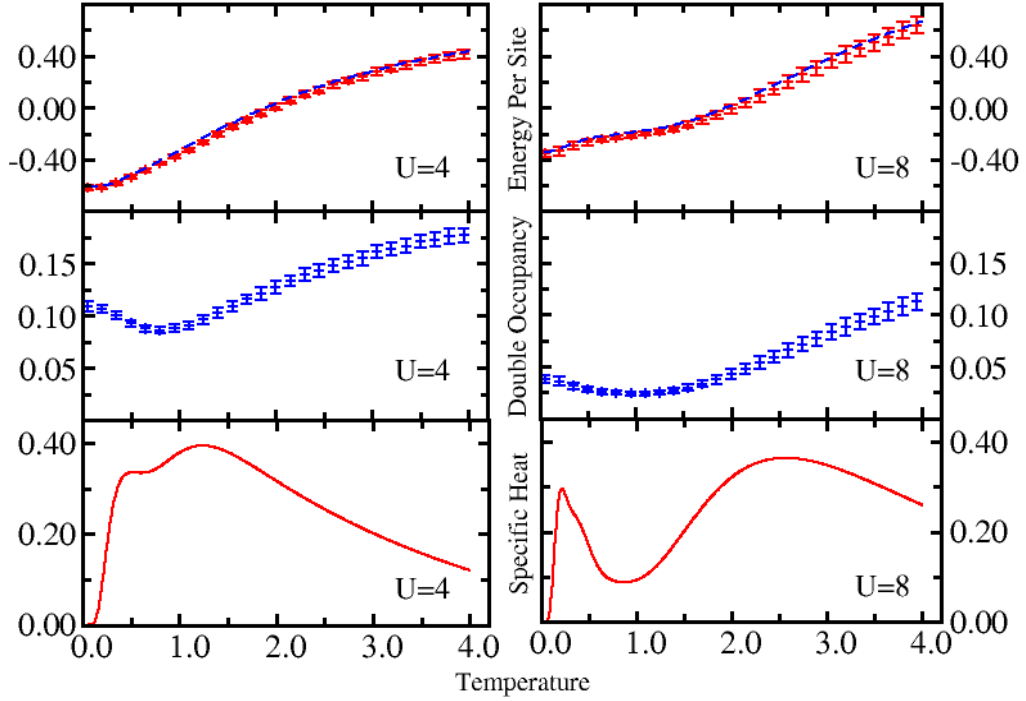

Figure 5. Energy (upper panels), double occupancy (middle panels), and specific heat (bottom panels) as a function of temperature for $U = 4$ (left panels) and $U = 8$ (right panels) of the system with six lattice sites. The QMC results (symbols) are in excellent agreement with the exact value (dashed lines) across the entire range of temperatures.

Fig. 6 shows the local moment ($L_{\alpha=0}$) and spin correlation functions ($L_{\alpha=1,2}$) as a function of temperature for $U = 4$ (left panel) and $U = 8$ (right panel) of the system with six lattice sites. With an increasing temperature, $L_0$ reaches its maximum at a certain temperature and then gradually decreases. The maximum value obtained for $L_0$ indicates that at this temperature, the degree of localization of electrons is the largest. The degree of delocalization of electrons reflects the formation of doublons; therefore, the trends of local moment and double occupancy are reversed, as shown in the upper panels of Fig. 6 and the middle panels of Fig. 5. From the lower panel of Fig. 6, one can see that $L_1$ is less than zero, which indicates an antiferromagnetic order at a finite temperature, which leads to the emergence of a specific heat peak at a lower temperature. As the temperature increases, $L_1$ decreases and tends to zero, reflecting a weakened antiferromagnetic order; this is in agreement with the exact results. In contrast, $L_2$ is greater than zero and gradually reduces to zero with a corresponding increase in temperature. It can be seen that, the error bar in $L_1$ and $L_2$ for U=8 is relatively large, which may be due to the limited simulation time, and/or intrinsic large fluctuations in spin correlation functions. Fortunately, the trends of $L_1$ and $L_2$ are in general consistent with the results of Shiba.[49,50]

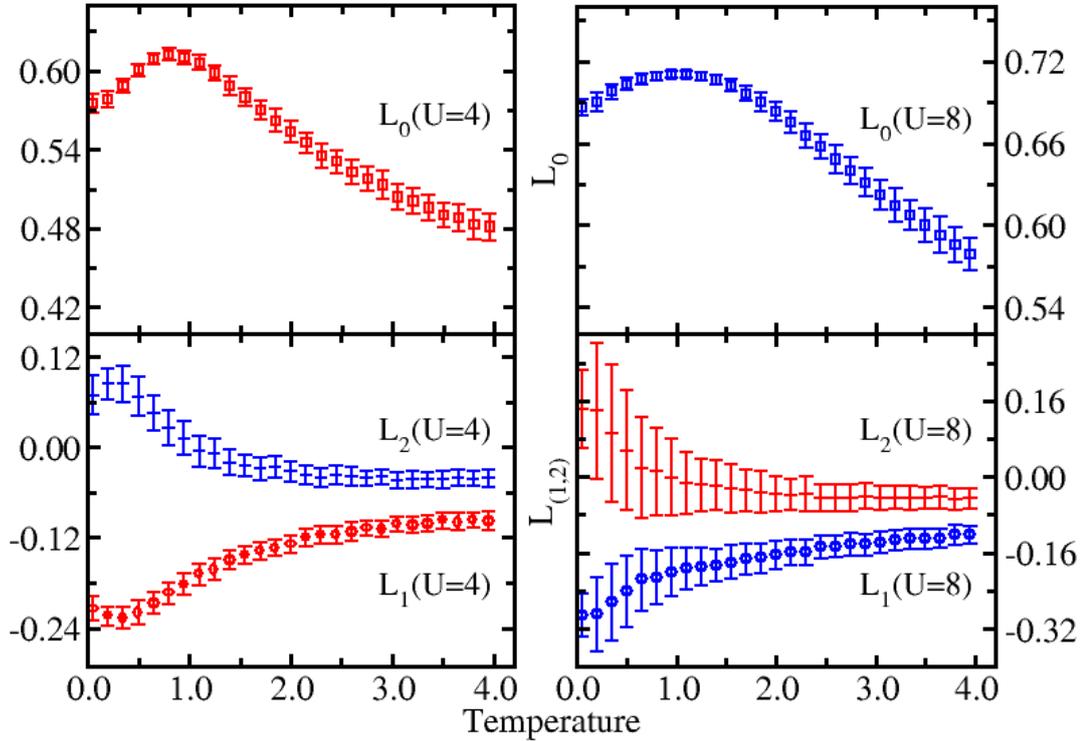

**Figure 6. Local moment ($L_{\alpha=0}$) and correlation functions ($L_{\alpha=1,2}$) as a function of temperature for $U = 4$ (left panel) and $U = 8$ (right panel) of the system with six lattice sites. $L_0$, $L_1$, and $L_2$ represent the local magnetic moment, the nearest-neighbor spin correlation, and the next-nearest-neighbor spin correlation, respectively.**

## V. Summary

We have proposed a path integral formula in field theory and a corresponding world-line quantum Monte Carlo algorithm. The remarkable feature of the current method is that neither determinants nor the Hubbard-Stratonovich Transformation is needed, which does strongly improve the accuracy and efficiency of Monte Carlo simulations. As an example, we have calculated the thermodynamic quantities and correlation functions of the one-dimensional Hubbard model at finite temperature. Our results are in excellent agreement with the exact values, confirming the reliability of our method. The most encouraging thing is that the computational cost has the square-law scaling with the size of systems. We believe that the current approach could be widely used in future.

**Acknowledgements:** Project supported by the National Natural Science Foundation of China (Grant No. 11874148). The computations were supported by ECNU Public Platform for Innovation.

**Author contributions statement**
D.Y. S conceived the theoretical idea and research. J. W. designed the Monte Carlo Code and carried out the theoretical calculations. J. W., W. P. and D.Y. S. analyzed the data and wrote the main manuscript text. All authors discussed the results and commented on the manuscript.

**Additional information**
Competing financial interests: The authors declare no competing financial interests.

**Data Availability Statement**
All relevant data are within the paper